# Plasmonic Light Trapping in an Ultrathin Photovoltaic Layer with Film-Coupled Metamaterial Structures

Hao Wang and Liping Wang [a)]

School for Engineering of Matter, Transport and Energy
Arizona State University, Tempe, Arizona, 85287 USA

## ABSTRACT

A film-coupled metamaterial structure is numerically investigated for enhancing the light absorption in an ultrathin photovoltaic layer of crystalline gallium arsenide (GaAs). The top subwavelength concave grating and the bottom metallic film could not only effectively trap light with the help of wave interference and magnetic resonance effects excited above the bandgap, but also practically serve as electrical contacts for photon-generated charge collection. The energy absorbed by the active layer is greatly enhanced with the help of the film-coupled metamaterial structure, resulting in significant improvement on the short-circuit current density by three times over a free-standing GaAs layer at the same thickness. The performance of the proposed light trapping structure is demonstrated to be little affected by the grating ridge width considering the geometric tolerance during fabrication. The optical absorption at oblique incidences also shows direction-insensitive behavior, which is highly desired for efficiently converting off-normal sunlight to electricity. The results would facilitate the development of next-generation ultrathin solar cells with lower cost and higher efficiency.

---
[a)] E-mail: liping.wang@asu.edu, Tel: 1-480-727-8615

Conventional solar cells are usually hundreds of microns in thickness due to the small absorption coefficient of semiconductor materials. Great efforts have been devoted to the investigation of thin-film solar cells with thickness of a few microns to reduce the cost for solar cells. However, effective light trapping is usually required to enhance light absorption in thin-film solar cells to achieve comparable or even better performance than conventional solar cells. Antireflection coatings[1,2] can enhance light absorption in solar cells at particular wavelengths due to the destructive interference between incident and reflected light. Surface texturing[3-5] is another approach to increase light absorption with multiple reflection inside the textured structure. Moreover, by introducing a back reflector[6], light absorption could also be enhanced by increasing the optical path length of light but subjected to the $4n^2$ limit.[7]

Plasmonic light trapping can achieve significant absorption enhancement in micro/nanostructured thin-film solar cells.[7-9] One dimensional (1D) back[10] and top[11,12] metallic gratings have been utilized to enhance the light absorption by exciting surface waves. To overcome the limitation of the polarization state with 1D gratings, 2D patch arrays have also been proposed for enhancing light trapping with polarization and directional independences.[13] Broadband absorption enhancement has been studied in grating structures with a plasmonic fractal.[14] Besides, plasmonic cavities in subwavelength hole arrays were also introduced for effective light trapping with 175% enhancement on power conversion efficiency.[15] In addition, scattering effect[16-19] and localized surface plasmon resonance[20,21] with nanoparticles were other plasmonic light trapping approaches. On the other hand, film-coupled metamaterials made of either convex[22,23] or concave gratings[24] on a dielectric spacer and a ground metal film have been recently studied for enhancing solar thermal energy harvesting. However, it is still a daunting challenge to effectively trap lights in ultrathin solar cells with thickness below 100 nm for enhanced light absorption and thereby solar-to-electricity conversion efficiency.

In this work, a film-coupled concave grating metamaterial structure is proposed to enhance light absorption in an ultrathin photovoltaic layer. As depicted in Fig. 1, a crystalline GaAs layer with thickness $t$ = 30 nm is sandwiched by a concave grating and substrate both made of Ag. The grating period is $\Lambda$ = 150 nm, while grating width and height are respectively $w$ = 30 nm and $h$ = 20 nm. The grating pattern from the proposed light trapping structure can be practically fabricated with electron-beam lithography, which has demonstrated nanostructures with feature size below 10 nm.[25] High-throughput nanoimprint lithography, which is capable for fabricating large-area nanostructures with feature size of 25 nm[26] or even down to 5 nm,[27] can be possibly employed to pattern the Ag concave grating with reduced cost. Recently, Kang et al. have successfully fabricated thin organic solar cells with nanoimprinted metallic electrodes with 70-nm linewidth and 40-nm thickness on a PEDOT:PSS layer of 50 nm.[28] A thin film of MgF$_2$ or transparent oxides can be deposited on the top to prevent the Ag concave grating from possible oxidization or chemical reaction in air under sunlight.[23] Note that, the top concave grating and the bottom metal film can not only facilitate plasmonic light trapping in the active layer, but also readily serve as the electrical contacts to harvest the photon-generated free charges from the ultrathin solar cell. Besides, the spreading resistance[29] between the front metallic contacts will be possibly reduced due to the small spacing distance (i.e. 120 nm) between metallic ridges.

The finite-difference-time-domain (FDTD) method (Lumerical Solutions, Inc.) was employed to perform the numerical simulation of the light-matter interaction with the proposed film-coupled metamaterial solar cell in a 0.15 μm × 0.15 μm × 2.5 μm domain with a manually refined mesh of 2.5 nm × 2.5 nm × 0.5 nm to ensure the numerical convergence. The optical properties of Ag and crystalline GaAs were taken from Palik.[30] A linearly-polarized plane wave source was placed above the structure to generate broadband incident light from 0.3 μm to 1 μm in wavelength with a spectral resolution of 1 nm. The plane of incidence is chosen to be the x-z plane, and the incidence angle $\theta$ is defined as the angle between incident wavevector $K$ and surface normal of the structure. A frequency-domain power monitor was located above the structure to calculate the spectral reflectance $R$. The spectral absorptance is obtained by $\alpha = 1-R-T$. Note that the structures with Ag substrate are opaque with zero transmittance $T$. Periodic boundary conditions were applied in x and y directions at normal incidence, while perfectly-matched layers with a reflection coefficient less than $10^{-6}$ were placed in z direction. Note that, the crystalline GaAs has a bandgap of 1.424 eV or equivalently 0.87 μm in wavelength.

Figure 2 shows the normal absorptance at transverse magnetic (TM) incidence with H field along y direction for the film-coupled metamaterial solar cell in comparison with a GaAs-on-Ag structure and a free-standing GaAs layer with the same thicknesses. For TM incidence, the H field is perpendicular to the plane of incidence (i.e., the x-z plane), while the E field is perpendicular to the plane of incidence for transverse electric (TE) waves. Note that the spectral response for the light trapping metamaterial is identical for TM and TE incidences due to the structural symmetry under normal incidence.

For a free-standing 30-nm GaAs layer, the absorptance decreases dramatically at longer wavelengths beyond 400 nm due to the low intrinsic absorption coefficient of GaAs. Therefore, it is highly desired that the light absorption could be significantly enhanced in this spectral region from 400 nm up to the bandgap of GaAs around 870 nm in wavelength, which is crucial to improve the electricity generation. When an Ag back reflector was placed below the ultrathin GaAs layer, which is the GaAs-on-Ag structure, it is observed that there exists an absorption peak at $\lambda = 0.72$ μm with absorptance $\alpha = 0.69$. This absorption peak is caused by the destructive interference between the incident and reflected waves inside the GaAs layer.[1] It is broader than that observed in other cavity resonators[31] due to the intrinsic loss of GaAs.

To better explain the resonant mechanism of the absorption peak at $\lambda = 0.72$ μm, the spectral normal reflectance of the GaAs-on-Ag structure can be analytically calculated by[1]:

$$R_\lambda = \left| \frac{r_{12} + r_{23}e^{2i\beta}}{1 + r_{12}r_{23}e^{2i\beta}} \right|^2 \tag{1}$$

where $r_{12}$ is the reflection coefficient from air to GaAs, and $r_{23}$ is the reflection coefficient from GaAs to Ag. Phase shift $\beta = 2\pi n t/\lambda$ considers the light propagation inside the ultrathin GaAs layer at normal incidence, where $n$ is the refractive index of GaAs. The absorptance of the GaAs-on-Ag structure is calculated from $1 - R_\lambda$, and the analytical thin-film optics method yields an absorption peak due to interference at $\lambda = 0.72$ μm with amplitude $\alpha = 0.66$, which is consistent with the FDTD simulation.

Now consider a subwavelength concave grating added onto the GaAs layer on Ag substrate, which is the proposed film-coupled light trapping structure. As shown in Fig. 2, the spectral absorptance exhibits two spectral peaks located at $\lambda = 0.67$ μm with unity absorptance and at $\lambda = 0.86$ μm with $\alpha = 0.9$, respectively. Thanks to these two absorption peaks, the film-coupled light trapping structure exhibits much greater light absorption compared with the free-standing GaAs film and the GaAs-on-Ag structure. Both absorption peaks are located above the bandgap of GaAs and thus could effectively enhance the light absorption for electron-hole pair generation. However, it is crucial to understand the physical mechanisms that are responsible for the enhanced light absorption in the film-coupled ultrathin solar cell structure.

The absorption peak at $\lambda_{FP} = 0.67$ μm is associated with the Fabry-Perot resonance in the Ag-GaAs-Ag cavity, which leads to near-unity absorptance. The presence of the additional top grating layer with 20-nm Ag modifies the reflection coefficient $r_{12}$ at the top interface of GaAs, which now becomes air-Ag-GaAs interface, resulting in the blue shift of the absorption peak from $\lambda_{FP} = 0.72$ μm to 0.67 μm with the addition of the top grating layer.

On the other hand, the long-wavelength resonance peak located at $\lambda_{MP} = 0.86$ μm is actually due to the excitation of magnetic resonance, or magnetic polariton (MP), which has been studied for tailoring optical properties of a similar film-coupled concave grating metamaterial structure but with an ultrathin lossless $SiO_2$ layer.[32] To help elucidate the physical mechanism, the electromagnetic field distribution at the resonance wavelength is plotted in Fig. 3, where arrows represent electric field vectors while the contour indicates magnetic field strength normalized to the incidence as $\log_{10}|H/H_0|^2$.

Figure 3(a) shows the x-z cross-section view of electromagnetic field distribution observed through the center plane of the cavity at y = Λ/2. Clearly, the electric field inside the GaAs layer forms a current loop between the top metallic ridge and bottom metal film along with localized magnetic field with 1 order of magnitude higher than the incidence, indicating strong light confinement at the resonance wavelength of $\lambda_{MP} = 0.86$ μm. This is exactly the electromagnetic field pattern when MP is excited,[22,32-34] in which charges oscillating along the metal surface around the dielectric layer form a resonant inductor-capacitor circuit, resulting in strong coupling between incident light and magnetic resonance inside the structure.

Figure 3(b) presents the x-y cross-section view of electromagnetic field distribution at the center plane of GaAs layer (i.e., z = 0). It can be seen that enhanced H field associated with MP is confined only under the ridge that is discontinuous in x direction with cavities at left and right sides. This can be explained by the fact that, under TM polarization with H field along y direction, oscillation of charges or current loops cannot be formed under the continuous metal ridges in x direction. On the other hand, charges could accumulate in discontinuous ridges terminated by the cavities, leading to excitation of MP.[32] Note that the resonance wavelength for MP is highly dependent on the geometric parameters such as the ridge width $w$. The proposed dimensions of the film-coupled metamaterial solar cell were carefully chosen in order to excite the MP resonance above the bandgap, thereby leading to potential high power generation due to enhanced light absorption.

Although light absorption can be significantly enhanced in the film-coupled concave grating structure by exciting MP and taking advantage of interference effect, the amount of energy absorbed by the photovoltaic layer, only which can contribute to the generation of electron-hole pairs, should be identified. Energy absorbed by other materials like metals in the structure essentially becomes loss. Therefore, it is crucial to evaluate the amount of energy absorbed by the active layer rather than that by the entire structure.

The energy absorbed per unit volume inside the GaAs layer can be obtained by:

$$P_{abs} = 0.5\varepsilon_0 \varepsilon'' \omega |E|^2 \tag{2}$$

while the absorbed power can be normalized to the incidence as:

$$\alpha_{GaAs} = \frac{\int P_{abs} dV}{0.5 c_0 \varepsilon_0 |E_{inc}|^2 A} \tag{3}$$

where $\varepsilon_0$ is the permittivity of vacuum, $\varepsilon''$ is the imaginary part for the relative permittivity of GaAs, $\omega$ is angular frequency, $E$ is electric field inside GaAs layer, $c_0$ is the speed of light in vacuum, $V$ is the volume of GaAs layer, $A$ is the area that the light source is incident upon, and $E_{inc}$ is the incident electric field. The normalized energy absorbed by the metals, or essentially loss, can be simply obtained by $\alpha - \alpha_{GaAs}$.

The absorbed energy normalized to incidence by the entire structure, GaAs, and metals is shown in Fig. 4(a) for the film-coupled light trapping structure. It can be observed that in the short-wavelength region from 0.3 μm to 0.45 μm, most of the energy are absorbed in the GaAs layer, while energy loss in Ag is negligible. Although the loss in Ag increases as wavelength increases, the energy absorbed by the GaAs could be as high as 70% at the absorption peak at $\lambda_{FP}$ = 0.67 μm due to wave interference effect. The GaAs layer could still absorb as much as 55% of the incident light close to its bandgap, thanks to the excitation of MP at $\lambda_{MP}$ = 0.86 μm, while another 34% is absorbed by the metals due to the ohmic loss when the induced electric current oscillates at the metal surfaces. Further efforts can be made to possibly reduce the energy loss by replacing metals with low-loss plasmonic materials in the visible and near-IR region such as transparent conducting oxides (e.g., ITO, AZO, GZO) and metal nitrides (e.g., TiN).[35]

To have a better idea on the effectiveness of light trapping with the film-coupled metamaterial structure, Fig. 4(b) compares the energy absorbed by the GaAs layer in the light trapping metamaterial structure compared with that in the free-standing GaAs layer and the GaAs-on-Ag structure. Clearly, the film-coupled structure has superior performance in trapping light over the free-standing photovoltaic layer that only absorbs 5% of light in the long wavelengths, thanks to the effects of MP and wave interference excited above the bandgap. Although the GaAs-on-Ag structure could effectively trap light with the interference effect, it is not practical as a front contact is always required to harvest free charges but might deteriorate the optical performance. The proposed film-coupled metamaterial structure could not only effectively trap light to enhance the light absorption but also readily serve as electrical contacts for practical considerations.

In order to quantitatively evaluate performance for the film-coupled light trapping structure as an ultrathin solar cell, the short-circuit current density $J_{sc}$ is calculated by:

$$J_{sc} = \frac{q}{hc_0} \int_{300nm}^{870nm} a_{GaAs}(\lambda) I(\lambda) \lambda d\lambda \qquad (4)$$

where $I(\lambda)$ is the solar radiative heat flux at AM1.5 (global tilt),[36] $q$ is elementary charge, and $h$ is Planck's constant. The calculated $J_{sc}$ values are respectively 14.9 mA/cm², 12.5 mA/cm², and 3.8 mA/cm² for the film-coupled metamaterial solar cell, the GaAs-on-Ag structure, and the free-standing GaAs layer. Clearly, the short-circuit current is greatly enhanced by relatively almost three times with the film-coupled light trapping structure over the free-standing layer. Although it seems that the relative enhancement of 20% over the GaAs-on-Ag structure is not that significant, the film-coupled metamaterial structure is much more practical design as a solar cell. Note that, 100% internal quantum efficiency is assumed in the calculation of $J_{sc}$, because the charge transport mechanism in the ultrathin GaAs layer with sub-100 nm thickness, which is little understood and needs to be further studied, would be expected to be quite different from the bulk counterpart.

In practice, sample fabrication processes usually suffer from the manufacturing tolerance, which would unavoidably cause the variation of the geometric dimensions from the designed values. The slight geometric uncertainty from fabrication may influence the performance of the light trapping solar cell. Here, the effect of ridge width $w$ on the performance of the proposed film-coupled light trapping structure is investigated. Figure 5 shows the total absorptance of the light trapping structure with ridge width $w$ = 25 nm, 30 nm, and 35 nm (i.e., geometric tolerance of ±16.7%), while other geometric parameters remain the same. It can be observed that the absorption peak associated with Fabry-Perot resonance is little affected by the ridge width, and the resonance wavelength remains to be around $\lambda_{FP}$ = 0.67 μm. This is because the small change in grating ridge width will not affect the interference condition inside the Fabry-Perot cavity.

On the other hand, the other absorption peak due to magnetic resonance slightly shifts with varied ridge width, i.e., blue-shifting from $\lambda_{MP} = 0.86$ μm to 0.83 μm when the actual width is 5 nm less than the desired value of 30 nm, or red-shifting to 0.88 μm with a larger width $w = 35$ nm. The dependence of the magnetic resonance wavelength on the grating width has been discussed previously in a similar film-coupled concave grating structure but with lossless dielectric spacer made of $SiO_2$,[32] rather than a semiconductor layer (GaAs) in the present study. Even though the magnetic resonance wavelength slightly changes with the ridge width, it turns out that, the short-circuit current $J_{sc}$ becomes 15.4 mA/cm$^2$ and 14.4 mA/cm$^2$ respectively with the ridge width of 25 nm and 35 nm, in comparison to the 14.9 mA/cm$^2$ with $w = 30$ nm. Therefore, a geometric tolerance of ±16.7% in the ridge width would only lead to a small relative error of ±3.4% in the short-circuit current of the proposed light trapping structure.

Finally, the optical absorption at oblique incidence is explored in order to examine the directional-sensitivity for the film-coupled light trapping structure. Figure 6 plots the total absorptance with varied incidence angles from 0° to 85° with a resolution of 5° at two selected wavelengths $\lambda_{FP} = 0.67$ μm and $\lambda_{MP} = 0.86$ μm, at which Fabry-Perot and magnetic resonances occur. Note that TM and TE incidences are no longer identical at oblique incidences, so the behavior at oblique incidences is studied separately at each polarization. Figure 6(a) shows the absorptance with TM-polarized incidence, in which it can be observed that at $\lambda_{FP} = 0.67$ μm, the absorptance barely changes with incidence angles $\theta < 30°$. When $\theta$ is further increased to 60°, the absorptance still remains as high as 0.9 or so. The unusual directional-insensitive absorption associated with Fabry-Perot cavity resonance is due to the lossy GaAs, which leads to a broad absorption peak in comparison with the sharp and direction-sensitive peaks with lossless dielectric cavies.[31] On the other hand, at $\lambda_{MP}$ = 0.86 µm, the absorption peak increases up to unity at the incidence angle of 65°. This is because the resonance strength for MP remains strong at TM oblique incidence, for which the incidence H field is always parallel to grating groove regardless of incidence angles.[32]

Figure 6(b) illustrates the total absorptance as a function of incidence angle for TE-polarized waves. At $\lambda_{FP}$ = 0.67 µm, the total absorptance has similar behavior as TM incidence, demonstrating highly insensitive behavior at oblique incidence angles. At $\lambda_{MP}$ = 0.86 µm, the absorptance remains around 0.85 up to $\theta$ = 30° but drops to 0.65 at $\theta$ = 60°. Note that the MP resonance strength will decrease at oblique TE incidence, as the H field parallel to grating groove will decrease at large incidence angles with TE polarization. The angle-resolved absorptance spectra at both polarizations clearly show the directional-insensitive behavior of the proposed light trapping structure, which is favorable for converting off-normal sunlight to electricity.

In conclusion, we have numerically demonstrated the plasmonic light trapping inside an ultrathin photovoltaic layer in the film-coupled metamaterial structure, which could not only readily serve as electrical contacts for charge harvesting but also effectively trap light with the help of interference and magnetic resonance effects above the bandgap, potentially leading to improved solar-to-electricity conversion efficiency. The short-circuit current density with the film-coupled metamaterial solar cell is enhanced by three times of that from a free-standing GaAs layer. The small variation on the grating ridge width due to fabrication tolerance has little effect on the performance of the proposed light trapping solar cell structure, whose optical absorption is also shown to be insensitive with the oblique incidences. The fundamental understanding gained here will facilitate the development of next-generation, ultrathin, low-cost, highly-efficient solar cells.

**Acknowledgment**

The authors thank the support from the US-Australia Solar Energy Collaboration - Micro Urban Solar Integrated Concentrators (MUSIC) project sponsored by the Australian Renewable Energy Agency (ARENA), as well as the ASU New Faculty Startup fund and Seed Project fund.

**Figure Captions:**

Fig. 1   Schematic of an ultrathin solar cell made of film-coupled concave grating metamaterial structure with dimensions: period $\Lambda = 150$ nm, width $w = 30$ nm, grating thickness $h = 20$ nm, and GaAs thickness $t = 30$ nm.

Fig. 2   Spectral absorptance at normal incidence of the film-coupled light trapping structure, in comparison with a free-standing GaAs film and a GaAs-on-Ag structure with the same thickness of GaAs layer. Note that the bandgap for crystalline GaAs is at $\lambda = 0.87$ μm.

Fig. 3   Electromagnetic field distribution at the wavelength $\lambda = 0.86$ μm when MP resonance is excited inside the film-coupled metamaterial: (a) x-z cross-section view at center plane of cavity (i.e. y = $\Lambda/2$) and (b) x-y cross-section view at center plane of GaAs (i.e. z = 0).

Fig. 4   (a) Absorptance, or normalized energy absorption by the GaAs layer, metals, and entire structure for the film-coupled light trapping structure; (b) Normalized energy absorbed by GaAs in the film-coupled light trapping metamaterial structure in comparison with a free-standing GaAs film and a GaAs-on-Ag structure with the same GaAs thickness.

Fig. 5   Total absorptance of the film-coupled light trapping structure with varying grating ridge width values of 25 nm, 30 nm, and 35 nm.

Fig. 6   Total absorptance of the film-coupled light trapping structure as a function of oblique incidence angles at two resonance peaks $\lambda_{MP} = 0.86$ μm and $\lambda_{FP} = 0.67$ μm for (a) TM waves; and (b) TE waves.

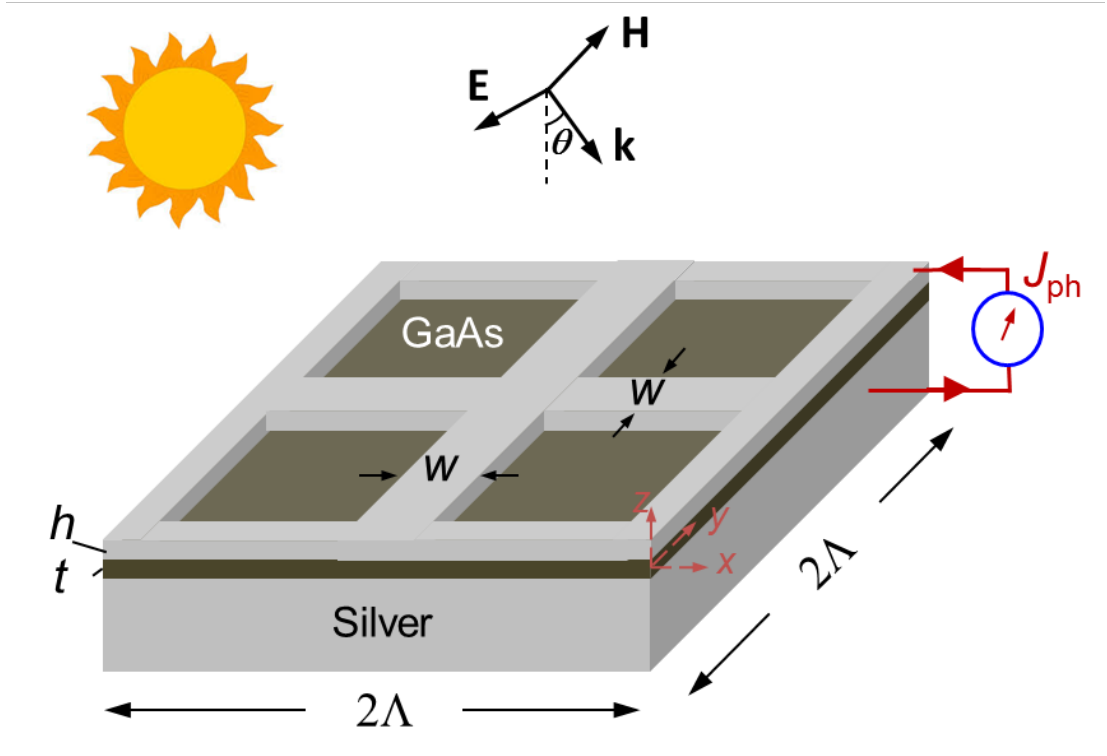

FIG. 1

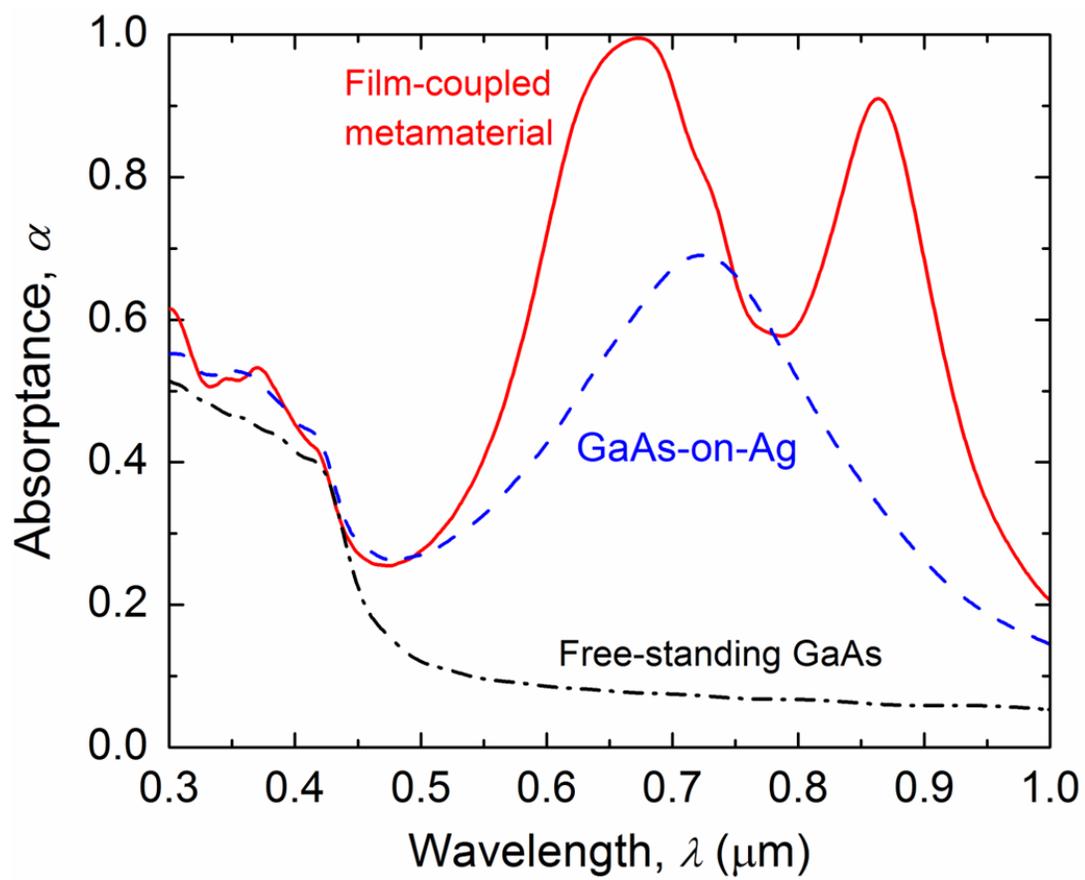

FIG. 2

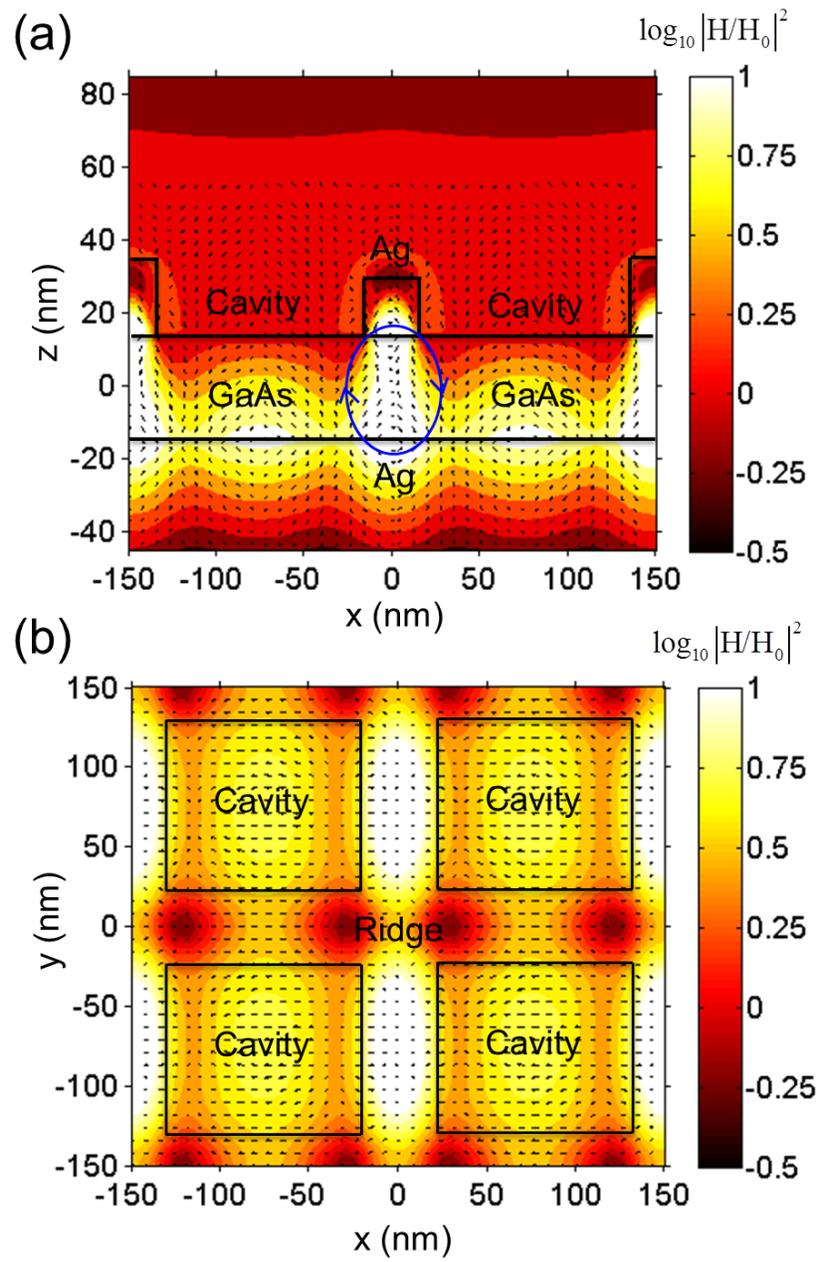

FIG. 3

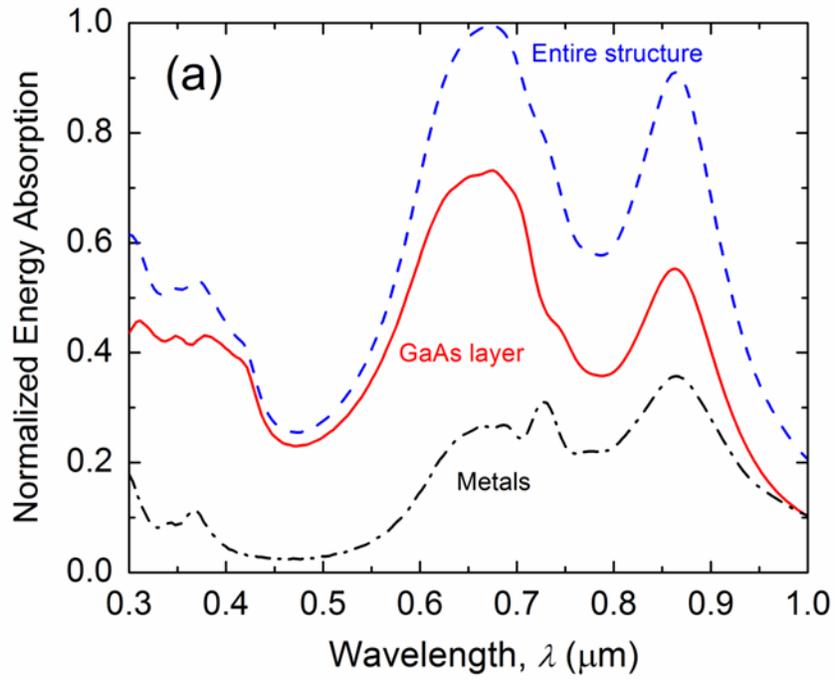

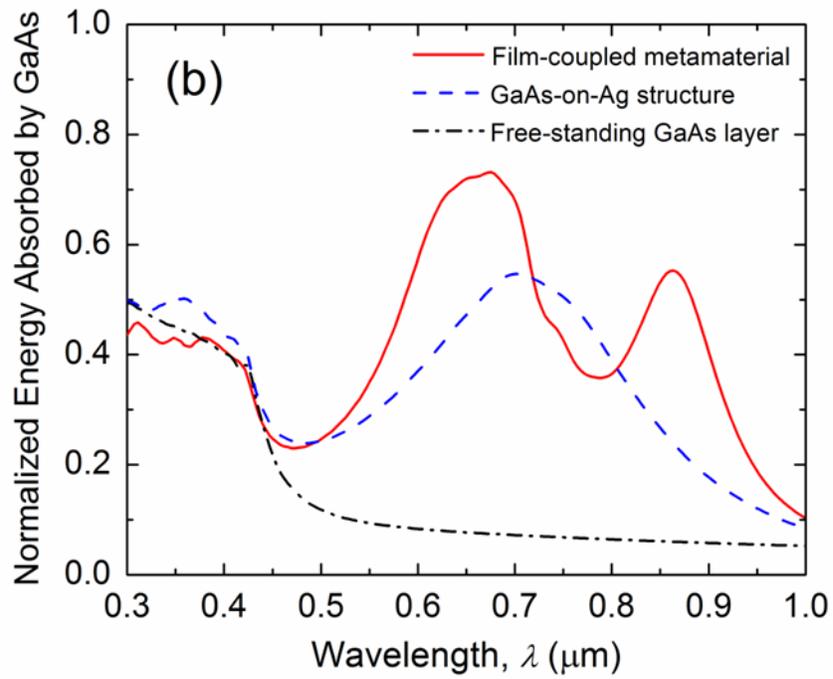

FIG. 4

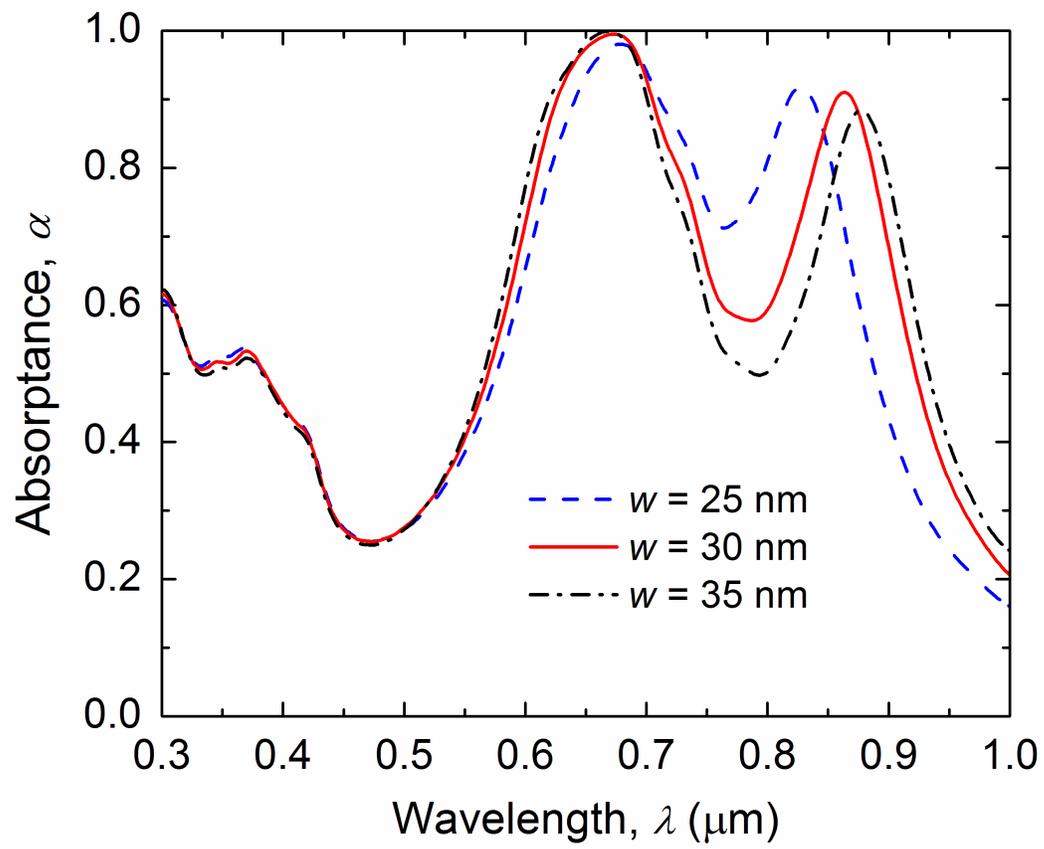

FIG. 5

## (a) TM waves

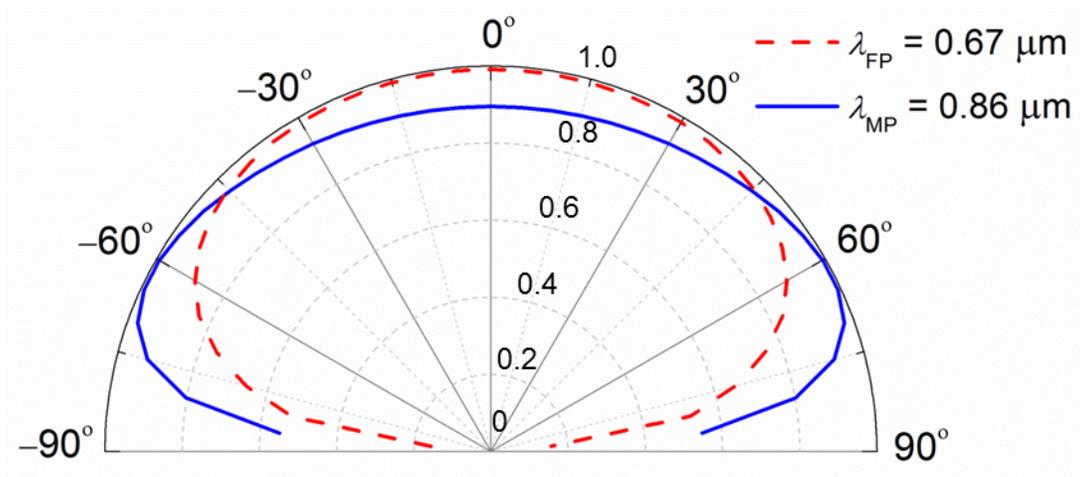

## (b) TE waves

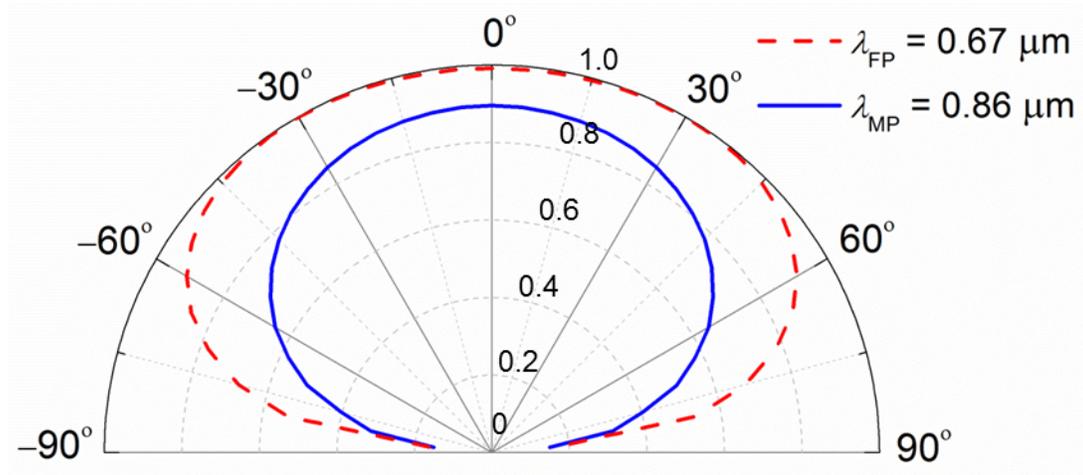

FIG. 6